\documentclass[aps,prb,twocolumn]{revtex4}
\usepackage{amssymb,amsbsy,graphicx,times}
\usepackage[latin1]{inputenc}

\begin{document}

\title{Critical current of a Josephson junction containing a conical magnet}
\author{G\'{a}bor B. Hal\'{a}sz}
\email{gh315@cam.ac.uk}
\address{Department of Material Science, University of Cambridge, Pembroke Street, Cambridge CB2 3QZ, UK }
\author{J. W. A. Robinson}
\address{Department of Material Science, University of Cambridge, Pembroke Street, Cambridge CB2 3QZ, UK }
\author{James F. Annett}
\address{H. H. Wills Physics Laboratory, University of Bristol, Royal Fort, Tyndall Avenue, Bristol BS8 1TL, UK}
\author{M. G. Blamire}
\address{Department of Material Science, University of Cambridge, Pembroke Street, Cambridge CB2 3QZ, UK}

\begin{abstract}
We calculate the critical current of a
superconductor/ferromagnetic/superconductor (S/FM/S) Josephson
junction in which the FM layer has a conical magnetic structure
composed of an in-plane rotating antiferromagnetic phase and an
out-of-plane ferromagnetic component. In view of the realistic
electronic properties and magnetic structures that can be formed
when conical magnets such as Ho are grown with a polycrystalline
structure in thin-film form by methods such as direct current
sputtering and evaporation, we have modeled this situation in the
dirty limit with a large magnetic coherence length ($\xi_f$). This
means that the electron mean free path is much smaller than the
normalized spiral length $\lambda/2\pi$ which in turn is much
smaller than $\xi_f$ (with $\lambda$ as the length a complete spiral
makes along the growth direction of the FM). In this physically
reasonable limit we have employed the linearized Usadel equations:
we find that the triplet correlations are short ranged and
manifested in the critical current as a rapid oscillation on the
scale of $\lambda/2\pi$. These rapid oscillations in the critical
current are superimposed on a slower oscillation which is related to
the singlet correlations. Both oscillations decay on the scale of
$\xi_f$. We derive an analytical solution and also describe a
computational method for obtaining the critical current as a
function of the conical magnetic layer thickness.
\end{abstract}



\maketitle


\section{Introduction}\label{intro}

The interaction of singlet-type superconductors (S) with
ferromagnetic materials in S/FM hybrid systems is a field of
extensive and ongoing research (see Refs.
\cite{ReviewBuzd2005,Bergeret2001,ReviewBerg2005} and references
therein). In proximity, the interaction of these competing electron
orders is characterized by an oscillating component in the Cooper
pair wave function which leads to a number of interesting phenomena:
the critical superconducting temperature $T_c$ dependence of S/FM
bilayers on FM layer thickness $d_{f}$,\cite{Tc oscillations in SF
bilayers1,Tc oscillations in SF bilayers2,Tc oscillations in SF
bilayers3,Tc oscillations in SF bilayers4} dependence of $T_c$ on
the orientation of FM layers in FM$'$/S/FM$''$ spin valves
\cite{F/S/F type memory devices1,F/S/F type memory devices2,F/S/F
type memory devices3,F/S/F type memory devices4,F/S/F type memory
devices5,F/S/F type memory devices6} and S/FM$'$/FM$''$ multilayers,
and finally the realization of $\pi$ coupling in S/FM/S Josephson
junctions. \cite{ZeroPi1,ZeroPi2,ZeroPi3,ZeroPi4,ZeroPi5}

The standard analysis of the S/FM systems has mostly assumed that
the FM is homogeneous and collinear, in which case only the singlet
superconducting correlation appears in the theory. Extending this
standard approach, theory strongly indicates that if the FM is
inhomogeneous and noncollinear, the longer-ranged triplet
superconducting correlations should then emerge at the S/FM
interface.\cite{ReviewBuzd2005,Bergeret2001,ReviewBerg2005} These
triplet correlations should then be insensitive to the exchange
field of the FM material and as such their proximity range is
expected to be similar to that of singlet pairs in a
superconductor/normal metal system.

Inhomogeneous magnetization exists in a range of material systems,
which can be classified into three categories: (1) magnetic domain
walls; (2) ferromagnetic multilayers such as when FM layers are
decoupled via a nonmagnetic (NM) spacer to form spin-active devices;
and (3) the intrinsically inhomogeneous and noncollinear magnetic
materials.

Domain walls were one of the first magnetically inhomogeneous
systems to be combined with superconductivity. Although experimental
studies of such systems are notoriously challenging because of the
need to control the magnetism at the nanometer scale, results and
analysis have indicated that domain walls are favorable nucleation
sites for
superconductivity.\cite{DomainWallSC1,DomainWallSC2,DomainWallSC3,DomainWallSC4,DomainWallSC5}
Theoretically, the emergence of triplet components in junctions
containing a single domain wall and or a multidomain ferromagnet
(MDFM) have been extensively
analyzed.\cite{SingleDomainWallJunctions1,SingleDomainWallJunctions2}
Recently, large area S/MDFM/S junctions have been
fabricated.\cite{TruptiPdNi2009} In this type of junction, the
amplitude of the critical current is expected to decay exponentially
with FM layer thickness. If singlet-type electron pairs are
scattered into triplet ones at the domain-wall regions, it is
expected that for a critical thickness of the MDFM the triplet
correlations will dominate over the singlet ones leading to slower
decay in the critical current with the MDFM thickness. So far,
evidence of a crossover from singlet to triplet-dominated transport
in these types of systems is nonexistent.

The second category, the ferromagnetic multilayers, have been
combined with superconductors in S/FM$'$/FM$'$/S junction form
junction form although most studies have been theoretical up to
now\cite{SFISF1,SFISF2,SFISF3,SFISF4} with only a few experiments
showing how the Josephson ground state is affected by the
orientation of the FM layers.\cite{SFISF5,SFISF6} The majority of
experimental studies have focused on how a superconducting layer is
modified by the relative orientation of the FMs. In these systems,
however, the triplet superconducting components that exist when the
FMs are noncollinear only transmit information about the direction
of the magnetic layers.\cite{Zimansky,Parity} To observe a
longer-ranged spin triplet proximity effect, it is thought that the
Josephson junction must contain three more
FMs,\cite{SFISF4,Houzet2007} with each offset from the other by an
angle $\theta \neq [0,\pi]$ (with $\pi$ as the antiparallel
configuration). In principle, the angle $\theta$ and thus the
triplet components could be controlled by the application of an
external magnetic field. Unfortunately, the implementation of a
large enough change in the angle $\theta$ with an applied magnetic
field is very difficult to realize without strongly suppressing the
superconductivity.

The third category, the intrinsically noncollinear magnets, is
potentially one of the simplest systems to combine with a
superconductor to experimentally study triplet
correlations.\cite{Linder2009} Recently,\cite{Sosnin2006}
interferometer measurements of superconducting Al coupled to the
rare-earth metal Ho have been made. In these Al/Ho/Al junctions,
superconducting phase periodic conductance oscillations were
observed indicating the presence of a longer-ranged proximity effect
when interpreted in the limit of a small coherence length in the Ho
relative to the length of a complete spiral
$\lambda$.\cite{Volkov2006} It is understood that the triplet
correlations were generated at the Al/Ho interface due to a rotating
magnetization present there and sustained by a continuous magnetic
spiral throughout the length of the Ho. A similar
explanation\cite{HalfMetal2,HalfMetal3,HalfMetal4,HalfMetal5} was
given for a long-ranged proximity effect observed in the half-metal
CrO$_2$.\cite{HalfMetal1} In this system, the triplet current was
shown to be insensitive to the strong polarization of the half
metal. Spin mixing at the interface is currently the best
explanation for the triplet proximity effect observed although a
better understanding of the interfaces that can exist in these types
of material systems is needed to verify this explanation. For Ho, it
is well known that growing it in thin-film form with a magnetic
spiral at the interface is difficult to achieve. This again
highlights a need to improve our understanding of the likely
properties and structures that can arise at the interface of
noncollinear magnets, such as Ho, with superconducting materials.

The magnetic structure\cite{Ho1,Ho2,Ho3} and electronic/thermal
properties\cite{Ho2} of the rare-earth Ho are well known. Its
magnetic structure has been characterized in bulk, single crystal,
and thin-film forms by neutron diffraction, x-ray diffraction, and
vibrating sample magnetometery. In thin-film form, the quality of
the conical magnetic structure is poorly understood although it is
well known that the growth method and growth conditions, crystal
forms, and interfacing materials affect the ordering range of the
magnetic structure.\cite{Ho1,Ho2}

Long-ranged magnetic ordering in Ho requires a coherent crystal
structure in which the $c$ axis is the screw axis with the moments
in the basal plane configured into a distorted helix parallel to the
$c$ axis. The quality of the Ho (e.g., impurity content and
roughness) and the strain at the NM/Ho interface are both important
factors in determining the scale of magnetic ordering; for example,
substantial intermixing at the NM/FM interface may disturb the
growth in the helix which may affect, smear out, or even destroy any
triplet correlations. Neutron-diffraction studies on epitaxial
(interfacially strained) Nb/Ho bilayer films grown by dc magnetron
sputtering\cite{Ho4} at high temperature suggest the presence of an
in-plane spiral (antiferromagnetic part) but no out-of-plane pitch
(ferromagnetic part) was detected even down to very low temperatures
$T \sim 1$ K. This implies that the strain at the Nb/Ho interface is
suppressing the ferromagnetic component. Further studies on
polycrystalline Nb/Ho/Nb trilayer films have also been
made.\cite{Ho5} In these films strain at the Nb/Ho interface is
lower and from Josephson-junction-type measurements a weakly conical
magnetic structure was confirmed from field-dependent measurements
of the junction's critical current as a function of the Ho spacer
layer thickness.

Altogether, the above review shows that although Ho can be grown on
top of thick Nb leads with a conical magnetic structure, strain at
the interface does weaken the ferromagnetic component possibly
implying a weaker magnetism. The motivation behind this work is to
complement the currently available theory on S/FM/S junctions with a
conical FM weak link by considering the physically reasonable
situation (see Sec. \ref{general}) in which the conical
ferromagnetic coherence length $\xi_f$ is much longer than the
normalized spiral length $\lambda/2\pi$. Because both $\xi_f$ and
$\lambda/2\pi$ are much larger than the electron mean free path
$\ell$ (dirty limit), the S/FM/S junction can be described within
the framework of the linearized Usadel equations.

This paper is organized as follows: Sec. \ref{magnetism} reviews the
magnetic and electronic properties of thin-film Ho and outlines the
important physical properties of the situation being analyzed in
this paper; in Secs. \ref{general} and \ref{josephson}, the general
theoretical framework in which we model a Josephson junction
containing a conical magnet is described with analytical solutions
to obtain the Josephson current explained; in Sec. \ref{numerical},
we present a computational method for calculating the Josephson
current, which is particularly useful to experimentalists.

\section{Magnetic structure and electronic properties of thin-film Holmium}\label{magnetism}

\begin{figure}[t!]
\centering
\includegraphics[width=8cm]{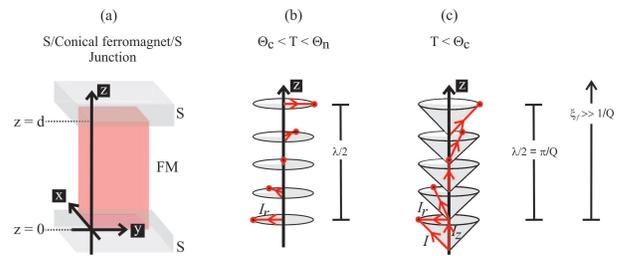}
\caption{(Color online) (a) An illustration of two singlet-type
superconductors (S) sandwiching a ferromagnet with a conical
magnetic structure (Ho). The two magnetic phases of Ho, (b) the
in-plane antiferromagnetic spiral phase of Ho below its N\'{e}el
temperature $\Theta_n$, and (c) the conical magnet phase of Ho below
its Curie temperature $\Theta_c$: the antiferromagnetic component
$I_r$ rotating in the $\{x,y\}$ plane; the ferromagnetic component
$I_z$ pitched towards the $z$ axis; and the resultant magnetization
vector $\mathbf{I}$ rotating on the surface of a cone. In the limit
considered in this paper, the ferromagnetic coherence length $\xi_f$
is much larger than $\lambda/2\pi$. \label{Figure1}}
\end{figure}

Consider the general structure of a conical magnet which consists of
a rotating in-plane magnetization and a constant out-of-plane
magnetization [see illustrations in Figs.
\ref{Figure1}(a)$-$\ref{Figure1}(c)]. The in-plane component is
effectively an antiferromagnetic (AFM) state which orders itself at
the N\'{e}el temperature $\Theta_n$, while the out-of-plane
component can be considered to be a ferromagnetic phase which orders
itself at the Curie temperature of the material. Thus, the strength
or exchange interaction energy $I$ of the ferromagnetic part is
related to the Curie temperature: $I \sim k_B \Theta_c$. The
in-plane component completes a full rotation in a distance of
$\lambda$ along the $z$ axis, which implies the distance along $z$
on which the in-plane component rotates in 1 rad is $\lambda/2\pi$.
From now on, $\lambda/2\pi$ will be referred to as the normalized
spiral length.

In the analysis that follows this section we shall assume that the
electron mean-free path $\ell$ is smaller than both the coherence
length of Ho $\xi_f$ and the normalized spiral length
$\lambda/2\pi$. For Ho in thin-film form, this limiting situation is
justified for the case when it is sputter deposited and
polycrystalline.

Polycrystalline thin films of Ho have a large residual resistivity
$\rho_0$ in the (6$-$12)$\times$10$^{-7}$ $\Omega$m range (see Refs.
\cite{Ho3} and \cite{Ho5} and references therein). A rough estimate
of the electron mean free path for the conduction electrons around 4
K using the relation $\ell = v_F m / \rho_0 n e^2$, where $v_F$, the
Fermi velocity is 1.6$\times$10$^6$ m/s,\cite{Ho3} $m$ is the mass
of the electron, and $n$ is the number density of free electrons,
gives a (0.5$-$1.0) nm range, which is smaller than both $\xi_f
\sim$ (6$-$7) nm and $\lambda/2\pi \sim$ 1.1 nm.\cite{Ho5} Even in
single-crystal form, the resistivity of thin-film Ho is large and
around 6$\times$10$^{-7}$ $\Omega$m with an electron mean-free path
of $\sim$ 1.0 nm in the $c$-axis orientation.

\section{General solution in a conical ferromagnet}\label{general}

Let us consider a conical ferromagnet FM, the axis of which
coincides with the $z$ axis. The magnetization vector and hence the
exchange field $\mathbf{I}=(I_x,I_y,I_z)$ has a constant axial
component $I_z$ and a radial component $I_r$, which rotates in the
$\{x,y\}$ plane with wave vector $Q = 2\pi / \lambda$ as we move
along the $z$ direction (see Fig. \ref{Figure1}). The $x$ and $y$
components of $\mathbf{I}$ are therefore given by
\begin{equation}
I_x(z)=I_r \cos(Qz) \quad \textrm{and} \quad I_y(z)=I_r \sin(Qz).
\label{magnet}
\end{equation}
In this section we find the solutions for the anomalous Green's
function in the conical FM, considering in particular the case of
large $Q$. We assume that the dirty limit is fulfilled, which means
that the FM coherence length ($\xi_f = \sqrt{\hbar D_f / |
\mathbf{I}|}$ with $D_f$ being the diffusivity of the FM) and the
normalized spiral length $\lambda/2\pi$ are both much larger than
the electron mean free path, i.e. $\xi_f, \lambda/2\pi \gg \ell$. If
it is also assumed that the anomalous function is sufficiently small
in the FM (which is the case if the S/FM interfacial resistance is
large enough),\cite{ReviewBuzd2005} we can use the linearized Usadel
equation \cite{Houzet2007}
\begin{equation}
\hbar D_f \frac{d^2 \hat{F}(z)}{dz^2} - 2\hbar |\omega| \hat{F}(z) -
\mathrm{sgn}(\omega) i [\mathbf{I}(z) \cdot \hat{\bm{\sigma}},
\hat{F}(z)]_+ = 0, \label{usadel}
\end{equation}
where the anomalous Green's function
\begin{equation}
\hat{F}=f_0\hat{1}+\mathbf{f} \cdot \hat{\bm{\sigma}} \label{green}
\end{equation}
is a matrix in spin space with $\mathbf{f}=(f_x,f_y,f_z)$, and
$\hat{\bm{\sigma}}=(\hat{\sigma}_x,\hat{\sigma}_y,\hat{\sigma}_z)$
is a vector containing the Pauli matrices. The component $f_0$ is an
even, while $f_x$, $f_y$ and $f_z$ are odd functions of the
frequency $\omega$. The Matsubara frequencies are given by
$\omega=(2n+1)\pi k_B T/\hbar$ with $n=0,\pm1,\pm2,...$ at
temperature $T$ and $[a,b]_+=ab+ba$ is the anticommutator. If we
substitute expression (\ref{green}) into Eq. (\ref{usadel}) we
obtain a set of equations for the four components of the anomalous
function $\hat{F}$,
\begin{equation}
\frac{1}{2}\hbar D_f \frac{d^2 f_0(z)}{dz^2} - \hbar \omega f_0(z) -
i[\mathbf{I}(z) \cdot \mathbf{f}(z)] = 0, \label{usadel10}
\end{equation}
\begin{equation}
\frac{1}{2}\hbar D_f \frac{d^2 f_{x,y,z}(z)}{dz^2} - \hbar \omega
f_{x,y,z}(z) - i I_{x,y,z}(z) f_0(z) = 0. \label{usadel1xyz}
\end{equation}
Since the symmetric properties of the Usadel equations with respect
to $\omega$ are trivial, only the case of $\omega>0$ is treated from
now; we already omitted $\mathrm{sgn}(\omega)$ and used $\omega$
instead of $|\omega|$ in Eqs. (\ref{usadel10}) and
(\ref{usadel1xyz}). After putting expressions (\ref{magnet}) into
Eqs. (\ref{usadel10}) and (\ref{usadel1xyz}), they can be simplified
with the substitution $f_{\pm} = f_x \pm i f_y$, which yields
\begin{equation}
\frac{d^2 f_0}{dz^2} - 2k_\omega^2 f_0 - i(2k_z^2 f_z + k_r^2 (f_+
e^{-iQz} + f_- e^{iQz})) = 0, \label{usadel20}
\end{equation}
\begin{equation}
\frac{d^2 f_{\pm}}{dz^2} - 2k_\omega^2 f_{\pm} - 2i k_r^2 f_0 e^{\pm
iQz} = 0, \label{usadel2xy}
\end{equation}
\begin{equation}
\frac{d^2 f_z}{dz^2} - 2k_\omega^2 f_z - 2i k_z^2 f_0 = 0.
\label{usadel2z}
\end{equation}
The quantities $k_\omega = \sqrt{\omega/D_f}$ and $k_{z,r} =
\sqrt{I_{z,r}/\hbar D_f}$ were also introduced at this step.

By searching a solution for Eqs. (\ref{usadel20})$-$(\ref{usadel2z})
in the form of
\begin{equation}
f_{0,z} = A_{0,z}e^{Kz} \quad \textrm{and} \quad f_{\pm} = A_{\pm}
e^{Kz} e^{\pm iQz} \label{trial}
\end{equation}
with $A_{\pm} = A_+' \pm A_-'$, we obtain a set of algebraic
equations for the amplitudes $A_0$, $A_z$, $A_+'$ and $A_-'$:
\begin{equation}
(K^2-2k_\omega^2)A_0 - 2i k_z^2 A_z - 2i k_r^2 A_+' = 0,
\label{eigen1}
\end{equation}
\begin{equation}
-2i k_z^2 A_0 + (K^2-2k_\omega^2)A_z = 0, \label{eigen2}
\end{equation}
\begin{equation}
-2i k_r^2 A_0 + (K^2-Q^2-2k_\omega^2)A_+' + 2i KQ A_-' = 0,
\label{eigen3}
\end{equation}
\begin{equation}
2i KQ A_+' + (K^2-Q^2-2k_\omega^2)A_-' = 0. \label{eigen4}
\end{equation}
A nontrivial solution only exists for the amplitudes if the
determinant of the system is zero; this gives a fourth-order
equation for $K^2$,
\begin{eqnarray}
&& \big{[} (K^2-2k_\omega^2)^2 + 4k_z^4 \big{]} \big{[}
(K^2-Q^2-2k_\omega^2)^2 + 4K^2 Q^2 \big{]} \nonumber \\ \nonumber \\
&& + 4k_r^4 (K^2-2k_\omega^2)(K^2-Q^2-2k_\omega^2)= 0, \label{det}
\end{eqnarray}
which is equivalent to the similar equation obtained by Volkov
\emph{et al.} \cite{Volkov2006} In their paper, they considered the
limit in which $k_r,k_z \gg k_\omega,Q$, whereas we take the limit
of $Q \gg k_r,k_z,k_\omega$. This limit seems to be appropriate in
the case of Ho, which has a conical magnetic structure with $\lambda
\approx$ 6 nm and therefore $Q \approx$ 1 nm$^{-1}$. The exchange
energies $I_r$ and $I_z$ can be estimated from the AFM and FM
ordering temperatures; assuming a typical diffusivity $D_f \approx$
5$\times$10$^{-4}$ m$^{2}$s$^{-1}$ and a temperature $T \approx$ 4
K, we obtain $k_r\sim$ 0.2 nm$^{-1}$ and $k_z,k_\omega \sim$ 0.05
nm$^{-1}$, which are all much smaller than $Q$.

In order to make the approximations more transparent, we introduce
the dimensionless quantities
\begin{equation}
\lambda = \frac{K^2}{Q^2} \quad \textrm{and} \quad
\epsilon_{r,z,\omega} = \frac{k_{r,z,\omega}^2}{Q^2} \label{epsilon}
\end{equation}
with $\epsilon_{r,z,\omega} \ll 1$. Without assuming anything about
the relative values of these small numbers, we take the case in
which their respective leading terms are on the same order of
magnitude; our results are therefore applicable to the general case
and the particular cases can be obtained by taking appropriate
limits. It turns out that the leading terms in the small quantities
$\epsilon_{r,z,\omega}$ are on the orders of $\epsilon_r^2$,
$\epsilon_z$ and $\epsilon_\omega$, hence we assume for the
approximations that $\epsilon_r^2 \sim \epsilon_z \sim
\epsilon_\omega$. Two roots of $\lambda$ are on the order of 1; if
we neglect every term smaller than $\epsilon_r^2$ from Eq.
(\ref{det}), we obtain
\begin{equation}
(1+\lambda)^2 = \frac{4\epsilon_\omega}{\lambda}
(1+\lambda+2\lambda^2) + \frac{4\epsilon_r^2}{\lambda} (1-\lambda).
\label{approx11}
\end{equation}
Since the terms on the right side are $\ll 1$, the left side has to
be small, which is only possible if $\lambda \approx -1$. In this
case we can substitute $\lambda = -1$ on the right side, hence Eq.
(\ref{approx11}) reduces to
\begin{equation}
(1+\lambda)^2 + 8\epsilon_\omega + 8\epsilon_r^2 = 0,
\label{approx12}
\end{equation}
which gives $\lambda = -1 \pm i \sqrt{8\epsilon_\omega +
8\epsilon_r^2}$. If we only keep the roots of $K$ for which
$\mathrm{Re}(K)>0$ and still neglect the terms smaller than
$\epsilon_r^2$, we obtain
\begin{equation}
K_{1,2} = \pm iQ + Q \sqrt{2\epsilon_\omega + 2\epsilon_r^2}
\label{K12}
\end{equation}
for the first two eigenvalues $K$. These correspond to rapidly
oscillating solutions (together with the rotation of the
magnetization vector), which decay much more slowly in the negative
direction. Since $K$ only appears as $K^2$ in Eq. (\ref{det}), the
roots with $\mathrm{Re}(K)<0$ can all be paired up with their
respective opposites and give the same solutions decaying in the
positive direction. Expression (\ref{K12}) without the term
$2\epsilon_r^2$ is equivalent to the result obtained by Bergeret
\emph{et al.} \cite{Bergeret2001} for a spiral ferromagnet ($I_z=0$,
hence $\epsilon_z=0$).

The two remaining roots for $\lambda$ are on the order of
$\epsilon_r^2$; in this case we can treat $\lambda$ as being small
and hence neglect larger powers of it. However, we must keep terms
up to the order of $\epsilon_r^4$ in Eq. (\ref{det}) to obtain the
quadratic equation
\begin{equation}
\lambda^2 - 4(\epsilon_r^2+\epsilon_\omega)\lambda +
4(\epsilon_z^2+\epsilon_\omega^2+2\epsilon_\omega \epsilon_r^2) = 0,
\label{approx2}
\end{equation}
which yields $\lambda = 2\epsilon_\omega + 2\epsilon_r^2 \pm
2\sqrt{\epsilon_r^4-\epsilon_z^2}$. Note that the leading terms are
indeed on the orders of $\epsilon_r^2$, $\epsilon_z$ and
$\epsilon_\omega$, as stated above. Two more eigenvalues $K$ with
$\mathrm{Re}(K)>0$ are obtained,
\begin{equation}
K_{3,4} = Q \sqrt{2\epsilon_\omega + 2\epsilon_r^2 \pm
2\sqrt{\epsilon_r^4-\epsilon_z^2}}. \label{K34}
\end{equation}
The behaviour of these solutions depends on the relative values of
$\epsilon_z$ and $\epsilon_r^2$ and now we can consider the two
particular cases. If $\epsilon_z>\epsilon_r^2$, the roots $K_{3,4}$
are complex conjugates and the solutions (\ref{K34}) describe a
slowly decaying oscillation in the negative direction. If
$\epsilon_z<\epsilon_r^2$, the roots $K_{3,4}$ are real, which means
that $\hat{F}$ decays exponentially without oscillations. This case
corresponds to almost in-plane magnetization and contains the limit
of the spiral ferromagnet; expression (\ref{K34}) reduces to $K_3 =
Q \sqrt{2\epsilon_\omega + 4\epsilon_r^2}$ and $K_4 = Q
\sqrt{2\epsilon_\omega}$ if $\epsilon_z=0$. The solution
corresponding to $K_4$ has zero amplitude in any S/FM system, while
$K_3$ coincides with the value obtained by Bergeret \emph{et al.}
\cite{Bergeret2001}

After determining the eigenvalues $K$ we calculate the corresponding
eigenvectors, i.e. the relative amplitudes of the different
components $f_0$, $f_z$ and $f_{\pm}$ in each solution. Since the
roots $K_{1,2}$ given by Eq. (\ref{K12}) appear as a direct
consequence of the rotation of the magnetization vector, we expect
the components $f_{\pm}$ to dominate in the corresponding solutions,
and hence we choose $A_{1+}'=A_{2+}'=1$ [$A_{1+}'$ and $A_{2+}'$ are
the $A_+'$ amplitudes appearing in Eqs.
(\ref{eigen1})$-$(\ref{eigen4}) for the solutions corresponding to
the roots $K_1$ and $K_2$, respectively]. Equation (\ref{eigen2})
shows that $A_z \ll A_0$ in these cases, while $A_0 \ll A_+'=1$
according to Eq. (\ref{eigen1}). It is valid therefore to take $A_z
\approx 0$, then use Eqs. (\ref{eigen1}) and (\ref{eigen4}) together
with Eq. (\ref{K12}) to obtain $A_{1-}'=-1$, $A_{2-}'=1$ and
$A_{10}=A_{20}=-2i \epsilon_r$ in the leading approximation.

The solutions corresponding to the other two roots $K_{3,4}$
predominantly consist of the components $f_0$ and $f_z$, therefore
we choose $A_{30}=A_{40}=1$. Equations (\ref{eigen3}) and
(\ref{eigen4}) show that $A_-' \ll A_+' \ll A_0$ in these cases,
which implies that $A_-' \approx 0$. Keeping this in mind, we can
apply Eq. (\ref{eigen3}) to get $A_{3+}'=A_{4+}'=-2i \epsilon_r$,
and Eq. (\ref{eigen2}) with Eq. (\ref{K34}) to obtain
\begin{equation}
A_{3z}=\frac{i \epsilon_z}{\epsilon_r^2 +
\sqrt{\epsilon_r^4-\epsilon_z^2}} \,\,\, \textrm{and} \,\,\,
A_{4z}=\frac{i \epsilon_z}{\epsilon_r^2 -
\sqrt{\epsilon_r^4-\epsilon_z^2}}. \label{A34}
\end{equation}
We can again consider the two cases: if $\epsilon_z>\epsilon_r^2$,
$A_{3z}$ and $A_{4z}$ are complex numbers with unit modulus, and
$A_{4z} = -A_{3z}^*$. In particular, Eq. (\ref{A34}) reduces to
\begin{equation}
A_{3z} \approx 1 + i \frac{\epsilon_r^2}{\epsilon_z} \quad
\textrm{and} \quad A_{4z} \approx -1 + i
\frac{\epsilon_r^2}{\epsilon_z} \label{A34lim1}
\end{equation}
in the limit of $\epsilon_z \gg \epsilon_r^2$. If
$\epsilon_z<\epsilon_r^2$, $A_{3z}$ and $A_{4z}$ are purely
imaginary numbers with $|A_{3z}||A_{4z}|=1$; they are being
approximated as
\begin{equation}
A_{3z} \approx \frac{i\epsilon_z}{2\epsilon_r^2} \quad \textrm{and}
\quad A_{4z} \approx \frac{2i \epsilon_r^2}{\epsilon_z}
\label{A34lim2}
\end{equation}
if $\epsilon_z \ll \epsilon_r^2$. In the limiting case of the spiral
ferromagnet ($\epsilon_z \rightarrow 0$), $A_{3z} \rightarrow 0$ and
$|A_{4z}| \rightarrow \infty$. The latter means that the solution
for $\hat{F}$ corresponding to the root $K_4$ has only $f_z$
component (and no singlet $f_0$ component); its amplitude is
therefore zero, as already mentioned above.

If we take the eigenvalues $K$ with $\mathrm{Re}(K)<0$, the
corresponding eigenvectors are similar to those corresponding to
their opposites; Eqs. (\ref{eigen1})$-$(\ref{eigen4}) show that the
amplitudes $A_0$, $A_z$ and $A_+'$ remain the same if we multiply
$K$ by ($-$1), while $A_-'$ changes sign. According to $A_{\pm} =
A_+' \pm A_-'$, this means that $A_+$ and $A_-$ are exchanged.
Keeping this in mind, we can write down the general solution for the
linearized Usadel equation (\ref{usadel}) in a conical ferromagnet.
If the FM occupies a region of thickness $d$ in the $z$ direction
(more specifically, the range $0<z<d$), the general solution can be
written as
\begin{widetext}
\begin{equation}
\hat{F} = \sum_{n=1}^4 \Big{[} B_n e^{-K_n z} \big{(} A_{n0} \hat{1}
+ A_{nz} \hat{\sigma}_z + \frac{1}{2} \sum_{\pm} A_{n \mp} e^{\pm
iQz} (\hat{\sigma}_x \mp i \hat{\sigma}_y) \big{)} + C_n e^{-K_n
(d-z)} \big{(} A_{n0} \hat{1} + A_{nz} \hat{\sigma}_z + \frac{1}{2}
\sum_{\pm} A_{n \pm} e^{\pm iQz} (\hat{\sigma}_x \mp i
\hat{\sigma}_y) \big{)} \Big{]}, \label{solution}
\end{equation}
\end{widetext}
where the eigenvalues $K_n$ and the relative amplitudes $A_{n0}$,
$A_{nz}$ and $A_{n\pm} = A_{n+}' \pm A_{n-}'$ are given by the
expressions obtained in this section. Note that the amplitudes
$A_{n+}$ and $A_{n-}$ are exchanged in the terms corresponding to
the solutions with $\mathrm{Re}(K)<0$, as mentioned above. The
amplitudes $B_n$ and $C_n$ are determined by the boundary conditions
at $z=0$ and $z=d$; these are discussed in Sec. \ref{josephson}.

\section{Josephson current in the S/FM/S junction}\label{josephson}

If the regions with $z<0$ and $z>d$ on the two sides of the FM are
occupied by two identical half-infinite superconductors S, we obtain
a S/FM/S junction. The axis $z$ of the conical FM is perpendicular
to the S/FM interfaces (see Fig. \ref{Figure1}). The two
superconductors have a phase difference of $\Phi$ with respect to
each other, so the bulk pairing potentials in the left and the right
S are given by $\Delta e^{-i \Phi /2}$ and $\Delta e^{i \Phi /2}$
($\Delta \in \mathbb{R}$). The normal and the anomalous Green
functions are $\hat{G}_{L,R} = G_s \hat{1}$ and $\hat{F}_{L,R} = F_s
\hat{1} e^{\mp i \Phi /2}$ in the bulk of the left and right
superconductors, respectively, where $G_s = \hbar \omega /
\sqrt{\hbar^2 \omega^2 + \Delta^2}$ and $F_s = \Delta /
\sqrt{\hbar^2 \omega^2 + \Delta^2}$. The normal state conductivities
of the S and the FM are $\sigma_s$ and $\sigma_f$, while the
interfacial resistance per unit area between the S and the FM is
denoted by $R$. We introduce the dimensionless quantities
\begin{equation}
\gamma = \frac{\sigma_f \xi_s}{\sigma_s \xi_f} \quad \textrm{and}
\quad \gamma_B = \frac{R \sigma_f}{\xi_f}, \label{gamma}
\end{equation}
where $\xi_s = \sqrt{\hbar D_s / 2\pi k_B T}$ is the superconducting
(quasiparticle) coherence length of the S with $D_s$ being its
diffusivity. If the interfacial resistance is large enough, i.e.
$\gamma_b \gg \max(1,\gamma)$, we can use rigid boundary conditions
at the S/FM interface;\cite{ReviewBuzd2005} we assume that the
pairing potential and hence the Green's functions are the same at
the interface as in the bulk material. Furthermore, because of
$\gamma_b \gg 1$ the anomalous function $\hat{F}$ is sufficiently
small in the FM, which verifies using the linearized Usadel
equations in the FM (see Sec. \ref{general}).

\begin{figure}[t!]
\centering
\includegraphics[width=8cm]{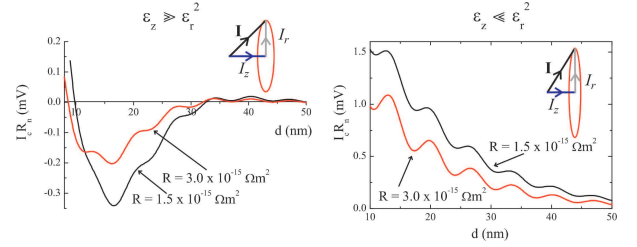}
\caption{(Color online) Typical $I_c R_n$ dependence on the
thickness $d$ of the FM layer ($\Delta =$ 2.2$\times$10$^{-22}$ J,
$Q =$ 9 $\times$10$^8$ m$^{-1}$, $D_f =$ 6 $\times$10$^{-4}$
m$^2$s$^{-1}$, $D_s =$ 2.5 $\times$10$^{-4}$ m$^2$s$^{-1}$,
$\sigma_f =$ 4 $\times$10$^6$ ($\Omega$m)$^{-1}$, $\sigma_s =$ 6
$\times$10$^6$ ($\Omega$m)$^{-1}$): (left) in the limit of
$\epsilon_z \gg \epsilon_r^2$ ($I_r/k_B=I_z/k_B=$ 100 K) and (right)
in the limit of $\epsilon_z \ll \epsilon_r^2$ ($I_r/k_B=$ 130 K,
$I_z/k_B<$ 4 K) for two different values of the interfacial
resistance $R$. \label{Figure2}}
\end{figure}

Assuming that $\gamma_b \gg \max(1,\gamma)$ is true, the rigid
boundary conditions are \cite{ReviewBuzd2005}
\begin{equation}
\hat{F}_L = G_s \hat{F}(0) - \gamma_B \xi_f \frac{d\hat{F}(0)}{dz}
\label{boundaryL}
\end{equation}
at the left side of the FM ($z=0$) and
\begin{equation}
\hat{F}_R = G_s \hat{F}(d) + \gamma_B \xi_f \frac{d\hat{F}(d)}{dz}
\label{boundaryR}
\end{equation}
at the right side of the FM ($z=d$). If the FM layer is thick enough
($d \gg \xi_f$), the terms containing $e^{-K_n (d-z)}$ can be
neglected from the general solution (\ref{solution}) near $z=0$,
while the terms containing $e^{-K_n z}$ can be neglected near $z=d$.
In this case we can take the components $f_0$, $f_z$ and $f_\pm$ of
Eq. (\ref{solution}) at $z=d$ and obtain equations for the
amplitudes $C_n$:
\begin{equation}
\frac{\Delta}{\hbar \omega} e^{i \Phi /2} = \sum_{n=1}^4 C_n A_{n0}
(1+\Gamma K_n), \label{boundary10}
\end{equation}
\begin{equation}
0 = \sum_{n=1}^4 C_n A_{nz} (1+\Gamma K_n), \label{boundary1z}
\end{equation}
\begin{equation}
0 = \sum_{n=1}^4 C_n A_{n \pm} (1+\Gamma(K_n \pm iQ)),
\label{boundary1xy}
\end{equation}
where the notation $\Gamma = \gamma_B \xi_f / G_s$ is used. Similar
equations hold for the amplitudes $B_n$ at $z=0$, the only
difference is that the sign of the phase $\Phi/2$ in the first term
of Eq. (\ref{boundary10}) is negative. Substituting $K_n$ and
$A_{n\pm}$ into Eq. (\ref{boundary1xy}) and using $|K_{3,4}| \ll Q$
yields
\begin{equation}
(C_{1,2} - i \epsilon_r (C_3+C_4)) + \Gamma (\pm Q \epsilon_r
(C_3+C_4) + k_0 C_{1,2}) = 0, \label{boundary2xy}
\end{equation}
where $k_0 = Q \sqrt{2\epsilon_\omega + 2\epsilon_r^2}$. Even though
$\Gamma Q \gg 1$, $\Gamma Q \epsilon_r \ll 1$ for realistic values
of the interfacial resistance; it follows from Eq.
(\ref{boundary2xy}) that $C_1,C_2 \ll C_3,C_4$, therefore the terms
containing $C_1$ and $C_2$ can be neglected from Eqs.
(\ref{boundary10}) and (\ref{boundary1z}). Those equations with
$A_{30} = A_{40} = 1$ take the form of
\begin{equation}
\frac{\Delta}{\hbar \omega} e^{i \Phi /2} = (C_3+C_4) + \Gamma (K_3
C_3 + K_4 C_4), \label{boundary20}
\end{equation}
\begin{equation}
0 = (A_{3z} C_3 + A_{4z} C_4) + \Gamma (A_{3z} K_3 C_3 + A_{4z} K_4
C_4). \label{boundary2z}
\end{equation}
In the following we work in the limit of $\epsilon_z \gg
\epsilon_r^2$, so we take $A_{3z}$ and $A_{4z}$ given by Eq.
(\ref{A34lim1}) to get
\begin{equation}
C_{3,4} = \frac{\Delta e^{i \Phi /2}}{2 \hbar \omega (1 +
K_{3,4}\Gamma)} \bigg{(} 1 \mp i \frac{\epsilon_r^2}{\epsilon_z}
\bigg{)}, \label{C34}
\end{equation}
where the expression (\ref{K34}) for the roots $K_{3,4}$ can be
approximated as $K_{3,4} \approx Q \sqrt{2\epsilon_\omega +
2\epsilon_r^2 \pm 2i \epsilon_z}$ in this case. Substituting $C_3$
and $C_4$ into Eq. (\ref{boundary2xy}) gives the remaining two
amplitudes,
\begin{equation}
C_{1,2} = \frac{\Delta e^{i \Phi /2} \epsilon_r (i \mp \Gamma Q)}{2
\hbar \omega (1 + k_0 \Gamma )} \bigg{(} \frac{1 - i \epsilon_r^2 /
\epsilon_z}{1 + K_3 \Gamma} + \frac{1 + i \epsilon_r^2 /
\epsilon_z}{1 + K_4 \Gamma} \bigg{)}. \label{C12}
\end{equation}
The term $i$ can be neglected from the numerator, because $\Gamma Q
\gg 1$. Furthermore, since $K_3$ and $K_4$ are complex conjugates if
$\epsilon_z>\epsilon_r^2$, the sum in Eq. (\ref{C12}) can be
simplified to
\begin{equation}
C_{1,2} = \mp \frac{\Delta e^{i \Phi /2} \epsilon_r \Gamma Q} {\hbar
\omega (1 + k_0 \Gamma )} \mathrm{Re} \bigg{(} \frac{1 - i
\epsilon_r^2 / \epsilon_z}{1+K_3 \Gamma} \bigg{)}. \label{C12lim}
\end{equation}
The results for $B_{3,4}$ and $B_{1,2}$ are the same as those given
by Eqs. (\ref{C34}) and (\ref{C12lim}) with the only difference
being a minus sign before $i \Phi / 2$ in the phase factor of
$\Delta$.

The Josephson current in the S/FM/S junction of area $A$ is given by
\cite{Houzet2007}
\begin{equation}
I = \frac{\pi \sigma_f A}{e} k_B T \sum_{\omega>0} \mathrm{Im}
\Big{[} \mathrm{Tr} \Big{(} \hat{F}^*(z) \hat{\sigma}_y
\frac{d\hat{F}(z)}{dz} \hat{\sigma}_y \Big{)} \Big{]},
\label{current}
\end{equation}
which can be evaluated in the range $0<z<d$. We can substitute the
solution (\ref{solution}) into Eq. (\ref{current}) and take $z=d$;
most terms do not give any imaginary contribution to the trace,
hence we obtain
\begin{equation}
I = \frac{\pi \sigma_f A}{e} k_B T \sum_{\omega>0} \mathrm{Im} (S_1
+ S_2), \label{current1}
\end{equation}
\begin{eqnarray}
S_1 & = & 2 \sum_{0,z}\Big{(} \sum_{n=1}^4 B_n^* A_{n(0,z)}^*
e^{-K_n^*d} \sum_{n=1}^4 C_n A_{n(0,z)} K_n \nonumber \\
& - & \sum_{n=1}^4 C_n^* A_{n(0,z)}^* \sum_{n=1}^4 B_n A_{n(0,z)}
K_n e^{-K_n d} \Big{)}, \label{S11}
\end{eqnarray}
\begin{eqnarray}
S_2 & = & - \sum_{\pm}\Big{(} \sum_{n=1}^4 B_n^* A_{n \mp}^*
e^{-K_n^* d} \sum_{n=1}^4 C_n A_{n \pm} (K_n \pm iQ) \nonumber \\
& - & \sum_{n=1}^4 C_n^* A_{n \mp}^* \sum_{n=1}^4 B_n A_{n \pm} (K_n
\pm iQ) e^{-K_n d} \Big{)}. \label{S21}
\end{eqnarray}
The first term $S_1$ mainly contains the solutions corresponding to
the eigenvalues $K_{3,4}$, while the second term $S_2$ is mainly
contributed by the solutions corresponding to $K_{1,2}$. By using
the values of $A_{n0}$, $A_{nz}$ and $A_{n\pm}$ in the limit
$\epsilon_z \gg \epsilon_r^2$, the expressions for $S_1$ and $S_2$
become
\begin{eqnarray}
S_1 & = & 4 K_3 e^{-K_3 d} \bigg{(} 1 + i \frac {\epsilon_r^2}
{\epsilon_z} \bigg{)} (B_4^* C_3 - C_4^* B_3) \nonumber \\
& + & 4 K_4 e^{-K_4 d} \bigg{(} 1 - i \frac {\epsilon_r^2}
{\epsilon_z} \bigg{)} (B_3^* C_4 - C_3^* B_4), \label{S12}
\end{eqnarray}
\begin{equation}
S_2 = 4 k_0 \big{[} e^{-K_1 d} (C_2^* B_1 - B_2^* C_1) + e^{-K_2 d}
(C_1^* B_2 - B_1^* C_2) \big{]} \label{S22}
\end{equation}
in the main approximation. Here we neglected the terms containing
the small amplitudes $A_{10}$, $A_{20}$, $A_{3\pm}$, $A_{4\pm}$ and
used the fact that $|A_{3z}| = |A_{4z}| = 1$ if
$\epsilon_z>\epsilon_r^2$. By taking $K_2 = K_1^*$ and $K_4 = K_3^*$
into account, then putting the above obtained expressions for $B_n$
and $C_n$ into Eqs. (\ref{S12}) and (\ref{S22}) we obtain
\begin{equation}
S_1 = 4i \sin (\Phi) \frac{\Delta^2}{\hbar^2 \omega^2} \mathrm{Re}
\bigg{[} \frac{K_3 e^{-K_3 d}} {(1 + K_3 \Gamma)^2} \bigg{(} 1 - i
\frac {\epsilon_r^2} {\epsilon_z} \bigg{)} \bigg{]}, \label{S13}
\end{equation}
\begin{eqnarray}
S_2 & = & 16i \sin (\Phi) \frac{\Delta^2}{\hbar^2 \omega^2} \bigg{[}
\mathrm{Re} \bigg{(} \frac{1 - i \epsilon_r^2 / \epsilon_z}{1 + K_3
\Gamma} \bigg{)} \bigg{]}^2 \nonumber \\
&& \times \frac{\epsilon_r^2 Q^2 \Gamma^2 k_0 e^{-k_0 d}} {(1 + k_0
\Gamma)^2} \cos (Qd). \label{S23}
\end{eqnarray}
The Josephson current through the S/FM/S junction therefore obeys
the formula $I = I_c \sin (\Phi)$, where the critical current $I_c$
is given by
\begin{widetext}
\begin{equation}
I_c R_n = 4 \pi k_B T \frac{d + 2 \xi_f \gamma_B}{e} \sum_{\omega>0}
\frac{\Delta^2}{\hbar^2 \omega^2} \Bigg{[} \mathrm{Re} \bigg{[}
\frac{K_3 e^{-K_3 d}} {(1 + K_3 \Gamma)^2} \bigg{(} 1 - i \frac
{\epsilon_r^2} {\epsilon_z} \bigg{)} \bigg{]} + \bigg{[} \mathrm{Re}
\bigg{(} \frac{1 - i \epsilon_r^2 / \epsilon_z}{1 + K_3 \Gamma}
\bigg{)} \bigg{]}^2 \frac{4 \epsilon_r^2 Q^2 \Gamma^2 k_0 e^{-k_0
d}} {(1 + k_0 \Gamma)^2} \cos (Qd) \Bigg{]} \label{IcRn1}
\end{equation}
\end{widetext}
with $R_n = (d + 2 \xi_f \gamma_B)/\sigma_f A$ being the normal
state resistance of the junction.

The dependence of $I_c R_n$ on the thickness $d$ predicted by Eq.
(\ref{IcRn1}) is plotted in Fig. \ref{Figure2} (left) for two values
of the interfacial resistance $R$. Both curves show a small, rapid
oscillation superimposed on a large, slow oscillation; they both
decay on the scale of the slow oscillation. Comparison between the
two curves demonstrates that an increase in $R$ reduces the current,
but makes the rapid oscillations relatively more pronounced.

The first term in Eq. (\ref{IcRn1}) gives the slow oscillation,
which is mainly due to the ``short-range" singlet and triplet
components of the anomalous Green's function $\hat{F}$ (i.e. the
singlet component $f_0$ and the triplet component $f_z$ with zero
projection on the $z$ axis). Conversely, the second term in Eq.
(\ref{IcRn1}) corresponds to the rapid oscillation, which is related
to the ``long-range" triplet components (i.e. the triplet components
$f_{\pm}$ with projection $\pm1$ on the $z$ axis). Note that in our
case the terms ``short-range" and ``long-range" do not mean any
difference in the respective decaying lengths; they are only defined
like this to be consistent with the notions used in other papers.
\cite{Volkov2006,Houzet2007}

Unlike the slow oscillation which is also present in a system with a
homogeneous FM, the rapid oscillation appears as a direct
consequence of the inhomogeneous magnetization. This is shown
clearly by the coincidence of its oscillation period and the
magnetic spiral wavelength $\lambda$. The magnetization changes
quickly with respect to the FM coherence length $\xi_f$, which
explains why the amplitude of the rapid oscillation is small
compared to that of the slow oscillation.

By taking the limit of $I_r \rightarrow$ 0 (and hence $\epsilon_r
\rightarrow$ 0), we recover an S/FM/S junction with a homogeneous FM
of exchange energy $I_z$. In this limit, Eq. (\ref{IcRn1}) reduces
to
\begin{equation}
I_c R_n = 4 \pi k_B T \frac{d + 2 \xi_f \gamma_B}{e} \sum_{\omega>0}
\frac{\Delta^2}{\hbar^2 \omega^2} \mathrm{Re} \bigg{[} \frac{K_3
e^{-K_3 d}} {(1 + K_3 \Gamma)^2} \bigg{]} \label{IcRn1lim}
\end{equation}
with the root $K_3$ taking the form of
\begin{equation}
K_3 = \sqrt{2 k_\omega^2 + 2i k_z^2}. \label{K3lim}
\end{equation}
This is the standard formula for $I_c R_n$ in a Josephson junction
with a homogeneous FM weak link. \cite{ReviewBuzd2005}

The same result is obtained in the limit of $Q \rightarrow \infty$
because $\epsilon_r = k_r^2/Q^2 \rightarrow$ 0 in this case. The
amplitude of the rapid oscillation vanishes as $Q^{-2}$, and hence
we recover Eq. (\ref{IcRn1lim}). This is physically understandable;
as the FM coherence length becomes very much larger than the
characteristic spiral wavelength, the radial magnetization
``averages out'' on the scale of $\xi_f$, which means that the
situation is equivalent to $I_r \rightarrow$ 0.

Now we can return to Eqs. (\ref{boundary20}) and (\ref{boundary2z})
and take the opposite limit, i.e. where $\epsilon_z \ll
\epsilon_r^2$. In this case we use $A_{3z}$ and $A_{4z}$ given by
Eq. (\ref{A34lim2}) to obtain different values for the amplitudes
$B_n$ and $C_n$. However, the expression (\ref{current1}) with the
same terms $S_1$ and $S_2$ still holds for the Josephson current.
After substituting the new values of $B_n$ and $C_n$ into Eq.
(\ref{current1}) and taking approximations valid in the given limit,
we recover $I = I_c \sin (\Phi)$ and obtain
\begin{widetext}
\begin{equation}
I_c R_n = 4 \pi k_B T \frac{d + 2 \xi_f \gamma_B}{e} \sum_{\omega>0}
\frac{\Delta^2}{\hbar^2 \omega^2} \Bigg{[} \bigg{[} \frac{K_3
e^{-K_3 d}} {(1 + K_3 \Gamma)^2} - \frac{\epsilon_z^2}
{4\epsilon_r^4} \frac{K_4 e^{-K_4 d}} {(1 + K_4 \Gamma)^2} \bigg{]}
+ \frac{4 \epsilon_r^2 Q^2 \Gamma^2 k_0 e^{-k_0 d}} {(1 + k_0
\Gamma)^2 (1 + K_3 \Gamma)^2} \cos (Qd) \Bigg{]} \label{IcRn2}
\end{equation}
\end{widetext}
for the critical current. The roots $K_{3,4}$ given by Eq.
(\ref{K34}) can be approximated as $K_3 \approx Q
\sqrt{2\epsilon_\omega + 4\epsilon_r^2}$ and $K_4 \approx Q
\sqrt{2\epsilon_\omega}$ if $\epsilon_z \ll \epsilon_r^2$.

The $I_c R_n$ dependence on $d$ as given by Eq. (\ref{IcRn2}) is
represented in Fig. \ref{Figure2} (right). The rapid oscillation is
similar as in the limit of $\epsilon_z \gg \epsilon_r^2$, but the
slow oscillation is absent; the other component of $I_c$ decays
exponentially without oscillation. The rapid oscillation is still
related to the ``long-range" triplet components, whereas the
exponential decay is to the ``short-range" singlet and triplet
components.

In the case between the two limits (where $\epsilon_z \sim
\epsilon_r^2$), both expressions (\ref{IcRn1}) and (\ref{IcRn2}) are
applicable, but they are not as accurate as when they are used in
their respective limiting cases. Since the approximations leading to
Eq. (\ref{IcRn1}) are less sensitive than those required for Eq.
(\ref{IcRn2}), the former is preferred to be used in such a case.

\section{Computational method for calculating the Josephson
current}\label{numerical}

In this section we describe an alternative method for obtaining the
Josephson current in the S/FM/S junction; it requires computational
power and does not yield an analytical formula, but is exact within
the framework of the linearized Usadel equations. The basic steps
are the same as in the previous sections: we first solve the
linearized Usadel equation (\ref{usadel}) with the boundary
conditions [Eqs. (\ref{boundaryL}) and (\ref{boundaryR})], then
evaluate Eq. (\ref{current}) at a suitable location.

Let us introduce the formal vector
\begin{equation}
\mathbf{F}(z) = \Big{(} f_0, f_x, f_y, f_z, \frac{df_0}{dz},
\frac{df_x}{dz}, \frac{df_y}{dz}, \frac{df_z}{dz} \Big{)}
\label{vectorF}
\end{equation}
containing the components of $\hat{F}$ and their respective
derivatives with respect to $z$. We denote the value of this vector
$\mathbf{F}_L$ at the right side of the left S (at $z=0$) and
$\mathbf{F}_R$ at the left side of the right S (at $z=d$). These
consist of the components of $\hat{F}_L$ and $\hat{F}_R$,
respectively. By using the method described in the Appendix, we can
relate $\mathbf{F}_L$ and $\mathbf{F}_R$ through a matrix type
equation. Since we know the first four components of both
$\mathbf{F}_L$ and $\mathbf{F}_R$, we can use this equation to
obtain the remaining four components (the derivatives).

The Josephson current is evaluated with Eq. (\ref{current}) in the S
side of the left S/FM interface; since we calculate the current in
the S, we must substitute $\sigma_s$ instead of $\sigma_f$ in Eq.
(\ref{current}). By using $f_{xL} = f_{yL} = f_{zL} = 0$ the formula
simplifies to
\begin{equation}
I = \frac{2 \pi \sigma_s A}{e} k_B T \sum_{\omega>0} \mathrm{Im}
\bigg{(} f_{0L} \frac{df_{0L}}{dz} \bigg{)}, \label{current2}
\end{equation}
where $f_{0L} = \Delta e^{-i \Phi /2} / \sqrt{\Delta^2 + \hbar^2
\omega^2}$ and $df_{0L}/dz$ is calculated by the method described in
the Appendix. By setting the phase difference to $\Phi = \pi/2$ we
obtain
\begin{equation}
I_c R_n = 2 \pi k_B T \frac{\sigma_s (d + 2 \xi_f \gamma_B)}{e
\sigma_f} \sum_{\omega>0} \mathrm{Im} \bigg{(} f_{0L}
\frac{df_{0L}}{dz} \bigg{)}. \label{IcRn3}
\end{equation}
Note that the values of $f_{0L}$ and $df_{0L}/dz$ depend on the
phase difference $\Phi$, as well as on other parameters describing
the junction.

\begin{figure}[h]
\centering
\includegraphics[width=8cm]{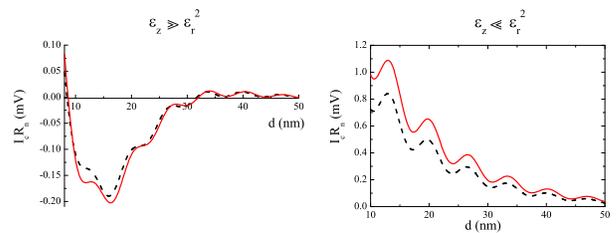}
\caption{(Color online) Comparison of the analytical (solid curve)
and computational (dashed curve) results for $I_c R_n$ in the
function of the thickness $d$: (left) in the limit of $\epsilon_z
\gg \epsilon_r^2$ and (right) in the limit of $\epsilon_z \ll
\epsilon_r^2$. The parameters used are identical to those in Fig.
\ref{Figure2} with $R=$ 3.0$\times$10$^{-15}$ $\Omega$ m$^2$.
\label{Figure3}}
\end{figure}

Evaluating $df_{0L}/dz$ requires inverting matrices and taking
matrix exponentials, therefore this method does not give an
analytical formula like expressions (\ref{IcRn1}) and (\ref{IcRn2}).
On the other hand, it gives the right result in the more general
case, even if $Q$ is not large [in which case neither Eq.
(\ref{IcRn1}), nor Eq. (\ref{IcRn2}) is applicable]. This method can
also be used to check analytical results; in our case it seems that
within their respective ranges, Eqs. (\ref{IcRn1}) and (\ref{IcRn2})
show good coincidence with the results obtained by the computational
method: see Fig. \ref{Figure3}.

\section{Summary}\label{conc}

We have calculated the Josephson current in a
superconductor/ferromagnetic/superconductor junction in which the
ferromagnet has a conical magnetic structure. In view of the
realistic interfaces that can exist between thin-film
superconductors (Al, Pb, and Nb being the elements typically used)
and thin films of conical magnets such as Ho, we have extended the
problem to a regime in which the ferromagnetic coherence length is
long compared to the electron mean-free path and the normalized
spiral length of the magnetic spiral.

From the materials point of view, the dirty-limit model we present
is physically reasonable and most applicable for when the Ho thin
film is polycrystalline. The electron mean-free path of thin film Ho
is $<$ 1.0 nm and its normalized spiral length is $\sim$ 1.1 nm,
whereas the coherence length in such films has recently been
determined to be in the 6$-$7 nm range.\cite{Ho5}

In this new situation, we have shown that the Josephson current is
highly sensitive to the length of the conical ferromagnet $\lambda$
with the current containing a rapidly oscillating component in the
function of the total conical magnetic thickness. These rapid
oscillations are superimposed on a much slower oscillation which has
a longer wavelength. The longer oscillation is directly linked to
the strength of the ferromagnetic component and mainly depends on
the singlet part of the anomalous Green's function $\hat{F}$. The
sign of the longer oscillation varies with multiple phase
transitions from $0$ to $\pi$ which depend on the thickness of the
magnetic layer. The rapid oscillations are linked to the triplet
components $f_{\pm}$ of the anomalous Green's function $\hat{F}$.
Although of a shorter wavelength to the slower oscillations, this
rapid oscillation also decays on the scale of the magnetic coherence
length.

The main feature of the results presented in this paper is that the
Josephson coupling through a conical ferromagnet may not be long
ranged as previously expected. In the limit considered, we have
shown that the proximity effect is short with a length scale
comparable to that of the proximity effects in a weak and collinear
ferromagnet. Thus, the theory explained in this paper is
complementary to previous
studies\cite{Linder2009,Sosnin2006,Volkov2006} which assume that the
magnetism of thin-film conical magnets is comparable to the
magnetism of conical magnets in bulk single-crystal form.

From the experimental view, the theory presented in this paper is
directly applicable to situations in which two singlet-type
superconductors are coupled via a rare-earth conical magnet (e.g.,
Ho). The experiment should be designed in such a way that the
current flowing through the
superconductor/ferromagnetic/superconductor junction is restricted
to the growth direction of the conical magnet, e.g., along the z
axis, such as in the case illustrated in Fig. \ref{Figure1}(a). For
similar experimental situations, see references \cite{ZeroPi5} and
 \cite{Bell2003}.


\begin{acknowledgments}

We are grateful to UK EPSRC for financial support (Grant No.
EP/F016611). J.W.A.R. thanks St John's College, Cambridge, and
G.B.H. acknowledges Prof. J. Driscoll of Trinity College, Cambridge,
for supporting his research.

\end{acknowledgments}


\appendix*

\section{Computational method in detail}\label{extra}

The boundary conditions at the S/FM interfaces are given by Eqs.
(\ref{boundaryL}) and (\ref{boundaryR}), and the equations relating
the appropriate derivatives,
\begin{equation}
G_s \frac{d\hat{F}_L}{dz} = \frac {\sigma_f} {\sigma_s}
\frac{d\hat{F}(0)}{dz} \quad \textrm{and} \quad \frac {\sigma_f}
{\sigma_s} \frac{d\hat{F}(d)}{dz} = G_s \frac{d\hat{F}_R}{dz}.
\label{boundaryD}
\end{equation}
If we assume rigid boundary conditions, $d\hat{F}_L/dz$ and
$d\hat{F}_R/dz$ are small because $\hat{F}_L, \hat{F}_R \approx$
constant. However, they are still not equal to zero and they must
not be neglected in this treatment. The boundary conditions
(\ref{boundaryL}), (\ref{boundaryR}) and (\ref{boundaryD}) can be
written as vector equations: $\mathbf{F}(0) = \mathbf{\hat{M}}_{fL}
\mathbf{F}_L$ and $\mathbf{F}_R = \mathbf{\hat{M}}_{Rf}
\mathbf{F}(d)$ with the 8$\times$8 matrices
\begin{equation}
\mathbf{\hat{M}}_{fL} =
\left(\begin{array}{cc} G_s^{-1} \hat{1} & R \sigma_s \hat{1} \\
\hat{0} & G_s \sigma_s \hat{1} / \sigma_f \end{array} \right),
\label{MfL}
\end{equation}
\begin{equation}
\mathbf{\hat{M}}_{Rf} = \left(\begin{array}{cc} G_s \hat{1} & R
\sigma_f \hat{1} \\ \hat{0} & G_s^{-1} \sigma_f \hat{1} / \sigma_s
\end{array} \right). \label{MRf}
\end{equation}
The symbols $\hat{1}$ and $\hat{0}$ denote the 4$\times$4 unit and
zero matrices, respectively. In order to deal with the interior of
the FM, we return to Eqs. (\ref{usadel20})$-$(\ref{usadel2z}) and
introduce new functions as $f_+ = f_+' e^{iQz}$ and $f_- = f_-'
e^{-iQz}$ (primes do not denote derivatives here). In this case we
obtain the equations
\begin{equation}
\frac{d^2 f_0}{dz^2} - 2k_\omega^2 f_0 - 2i k_z^2 f_z - i k_r^2
(f_+' + f_-') = 0, \label{usadel30}
\end{equation}
\begin{equation}
\frac{d^2 f_{\pm}'}{dz^2} \pm 2iQ \frac{df_{\pm}'}{dz} - Q^2
f_{\pm}' - 2k_\omega^2 f_{\pm}' - 2i k_r^2 f_0 = 0,
\label{usadel3xy}
\end{equation}
\begin{equation}
\frac{d^2 f_z}{dz^2} - 2k_\omega^2 f_z - 2i k_z^2 f_0 = 0.
\label{usadel3z}
\end{equation}
Equations (\ref{usadel30})$-$(\ref{usadel3z}) can be written
compactly as
\begin{equation}
\frac{d \mathbf{F}'(z)}{dz} = \mathbf{\hat{M}}_{F} \mathbf{F}'(z)
\quad \textrm{with} \quad
\mathbf{\hat{M}}_{F} = \left(\begin{array}{cc} \hat{0} & \hat{1} \\
\hat{K} & \hat{L}\end{array} \right) \label{MF}
\end{equation}
if we introduce the formal vector
\begin{equation}
\mathbf{F}'(z) = \Big{(} f_0, f_+', f_-', f_z, \frac{df_0}{dz},
\frac{df_+'}{dz}, \frac{df_-'}{dz}, \frac{df_z}{dz} \Big{)}.
\label{vectorFprime}
\end{equation}
The 4$\times$4 matrices $\hat{K}$ and $\hat{L}$ are given by
\begin{equation}
\hat{K} = \left(\begin{array}{cccc} 2k_\omega^2 & i k_r^2 & i k_r^2
& 2i k_z^2 \\ 2i k_r^2 & Q^2 + 2k_\omega^2 & 0 & 0
\\ 2i k_r^2 & 0 & Q^2 + 2k_\omega^2 & 0 \\ 2i k_z^2 & 0 & 0 &
2k_\omega^2 \end{array} \right), \label{matrixK}
\end{equation}
\begin{equation}
\hat{L} = \left(\begin{array}{cccc} 0 & 0 & 0 & 0 \\ 0 & -2iQ & 0 &
0 \\ 0 & 0 & 2iQ & 0 \\ 0 & 0 & 0 & 0 \end{array} \right),
\label{matrixL}
\end{equation} as it is clear from Eqs. (\ref{usadel30})$-$(\ref{usadel3z}).
Solving Eq. (\ref{MF}) in the region $0<z<d$ gives $\mathbf{F}'(d) =
\exp (d \mathbf{\hat{M}}_{F}) \mathbf{F}'(0)$. We also introduce the
conversion matrices
\begin{equation}
\mathbf{\hat{M}}_{Ff} = \left(\begin{array}{cccccccc} 1 & 0 & 0 & 0
& 0 & 0 & 0 & 0 \\ 0 & 1 & i & 0 & 0 & 0 & 0 & 0 \\ 0 & 1 & -i & 0 &
0 & 0 & 0 & 0 \\ 0 & 0 & 0 & 1 & 0 & 0 & 0 & 0 \\ 0 & 0 & 0 & 0 & 1
& 0 & 0 & 0 \\ 0 & -iQ & Q & 0 & 0 & 1 & i & 0 \\ 0 & iQ & Q & 0 & 0
& 1 & -i & 0 \\ 0 & 0 & 0 & 0 & 0 & 0 & 0 & 1
\end{array} \right), \label{MFf}
\end{equation}
\begin{equation}
\mathbf{\hat{M}}_{fF} = \left(\begin{array}{cccccccc} 1 & 0 & 0 & 0
& 0 & 0 & 0 & 0 \\ 0 & \alpha/2 & 1/2\alpha & 0 & 0 & 0 & 0 & 0
\\ 0 & -i\alpha/2 & i/2\alpha & 0 & 0 & 0 & 0 & 0 \\ 0 & 0 & 0 & 1 & 0 & 0
& 0 & 0 \\ 0 & 0 & 0 & 0 & 1 & 0 & 0 & 0 \\ 0 & i\alpha Q/2 &
-iQ/2\alpha & 0 & 0 & \alpha/2 & 1/2\alpha & 0 \\ 0 & \alpha Q/2 &
Q/2\alpha & 0 & 0 & -i\alpha/2 & i/2\alpha & 0 \\ 0 & 0 & 0 & 0 & 0
& 0 & 0 & 1 \end{array} \right), \label{MfF}
\end{equation}
where $\alpha = e^{iQd}$. These can be used to convert
$\mathbf{F}(0)$ to $\mathbf{F}'(0)$ as $\mathbf{F}'(0) =
\mathbf{\hat{M}}_{Ff} \mathbf{F}(0)$ and $\mathbf{F}'(d)$ to
$\mathbf{F}(d)$ as $\mathbf{F}(d) = \mathbf{\hat{M}}_{fF}
\mathbf{F}'(d)$. By taking the matrices defined in Eqs. (\ref{MfL}),
(\ref{MRf}), (\ref{MF}), (\ref{MFf}) and (\ref{MfF}) we obtain
\begin{equation}
\mathbf{F}_R = \mathbf{\hat{M}} \mathbf{F}_L, \label{vectoreq}
\end{equation}
\begin{equation} \mathbf{\hat{M}} =
\mathbf{\hat{M}}_{Rf} \mathbf{\hat{M}}_{fF} \exp (d
\mathbf{\hat{M}}_{F}) \mathbf{\hat{M}}_{Ff} \mathbf{\hat{M}}_{fL}.
\label{matrixM}
\end{equation}

Due to rigid boundary conditions, the components of $\hat{F}_L$ are
$f_{0L} = \Delta e^{-i \Phi /2} / \sqrt{\Delta^2 + \hbar^2
\omega^2}$ and $f_{xL} = f_{yL} = f_{zL} = 0$, while the components
of $\hat{F}_R$ are $f_{0R} = \Delta e^{i \Phi /2} / \sqrt{\Delta^2 +
\hbar^2 \omega^2}$ and $f_{xR} = f_{yR} = f_{zR} = 0$. The first
four components of the vectors $\mathbf{F}_L$ and $\mathbf{F}_R$ are
therefore known and Eq. (\ref{vectoreq}) can be used to obtain the
remaining four components (the derivatives). If we divide the matrix
$\mathbf{\hat{M}}$ into 4$\times$4 blocks as
\begin{equation}
\mathbf{\hat{M}} = \left(\begin{array}{cc} \hat{M}_{11} &
\hat{M}_{12} \\ \hat{M}_{21} & \hat{M}_{22}
\end{array} \right), \label{matrixMblock}
\end{equation}
and then take the first four components of Eq. (\ref{vectoreq}), a
pre-multiplication by $\hat{M}_{12}^{-1}$ gives the derivatives as
\begin{equation}
\left(\begin{array}{c} df_{0L}/dz \\ df_{xL}/dz \\
df_{yL}/dz \\ df_{zL}/dz \end{array} \right) = \hat{M}_{12}^{-1}
\Bigg{[} \left(\begin{array}{c} f_{0R} \\ f_{xR} \\ f_{yR} \\ f_{zR}
\end{array}
\right) - \hat{M}_{11} \left(\begin{array}{c} f_{0L} \\ f_{xL} \\ f_{yL} \\
f_{zL} \end{array} \right) \Bigg{]}. \label{vectoreq1}
\end{equation}
The first component $df_{0L}/dz$ is used in Eqs. (\ref{current2})
and (\ref{IcRn3}).


\begin{references}
\bibitem{ReviewBuzd2005} A. I. Buzdin, Rev. Mod. Phys. \textbf{77}, 935 (2005).
\bibitem{Bergeret2001} F. S. Bergeret, A. F. Volkov and K. B. Efetov, Phys. Rev. B \textbf{64}, 134506 (2001).
\bibitem{ReviewBerg2005} F. S. Bergeret, A. F. Volkov and K. B. Efetov, Rev. Mod. Phys. \textbf{77}, 1321 (2005).
\bibitem{Tc oscillations in SF bilayers1} J. S. Jiang, D. Davidovi\'{c}, Daniel H. Reich, and C. L. Chien, Phys. Rev. Lett. \textbf{74}, 314 (1995).
\bibitem{Tc oscillations in SF bilayers2} Th. M\"{u}hge, N. N. Garif'yanov, Yu. V. Goryunov, G. G. Khaliullin, L. R. Tagirov, K. Westerholt, I. A. Garifullin, and H. Zabel, Phys. Rev. Lett. \textbf{77}, 1857 (1996).
\bibitem{Tc oscillations in SF bilayers3} M. V\'{e}lez, M. C. Cyrille, S. Kim, J. L. Vicent, Ivan K. Schuller, Phys. Rev. B \textbf{59}, 14659 (1999).
\bibitem{Tc oscillations in SF bilayers4} P. Koorevaar, Y. Suzuki, R. Coehoorn, J. Aarts, Phys. Rev. B \textbf{49}, 441 (1994).
\bibitem{F/S/F type memory devices1} V. Pe\H{n}a, Z. Sefrioui, D. Arias, C. Leon, J. Santamaria, J. L. Martinez, S. G. E. te Velthuis, and A. Hoffmann, Phys. Rev. Lett. \textbf{94}, 057002 (2005).
\bibitem{F/S/F type memory devices2} I. C. Moraru, W. P. Pratt, Jr., and N. O. Birge, Phys. Rev. Lett. \textbf{96}, 037004 (2006).
\bibitem{F/S/F type memory devices3} A. Yu. Rusanov, S. Habraken, and J. Aarts, Phys. Rev. B \textbf{73}, 060505(R) (2006).
\bibitem{F/S/F type memory devices4} J. Y. Gu, C.-Y. You, J. S. Jiang, J. Pearson, Ya. B. Bazaliy, and S. D. Bader, Phys. Rev. Lett. \textbf{89}, 267001 (2002).
\bibitem{F/S/F type memory devices5} A. Potenza and C. H. Marrows, Phys. Rev. B \textbf{71}, 180503(R) (2005).
\bibitem{F/S/F type memory devices6} D. Stamopoulos, E. Manios, and M. Pissas, Phys. Rev. B \textbf{75}, 014501 (2007).
\bibitem{ZeroPi1} V. V. Ryazanov, V. A. Oboznov, A. Yu. Rusanov, A. V. Veretennikov, A. A. Golubov, and J. Aarts, Phys. Rev. Lett. \textbf{86}, 2427 (2001).
\bibitem{ZeroPi2}T. Kontos, M. Aprili, J. Lesueur, F. Gen\^{e}t, B. Stephanidis, and R. Boursier, Phys. Rev. Lett. \textbf{89}, 137007 (2002).
\bibitem{ZeroPi3} C. Bell1, R. Loloee, G. Burnell, and M. G. Blamire, Phys. Rev.B \textbf{71}, 180501 (2005).
\bibitem{ZeroPi4} V. A. Oboznov, V. V. Bol'ginov, A. K. Feofanov, V. V. Ryazanov, and A. I. Buzdin, Phys. Rev. Lett. \textbf{96}, 197003 (2006).
\bibitem{ZeroPi5} J. W. A. Robinson, S. Piano, G. Burnell, C. Bell, and M. G. Blamire, Phys. Rev. Lett. \textbf{97}, 177003 (2006).
\bibitem{DomainWallSC1}  F.S. Bergeret, A.F. Volkov, and K. B. Efetov, Phys. Rev. Lett. \textbf{86}, 4096 (2001)
\bibitem{DomainWallSC2} A. Yu. Aladyshkin, A. I. Buzdin, A. A. Fraerman, A. S. Melnikov, D. A. Ryzhov, and A. V. Sokolov, Phys. Rev. B \textbf{68}, 184508 (2003).
\bibitem{DomainWallSC3} Z. Yang, M. Lange, A. Volodin, R. Szymczak and V. V. Moshchalkov, Nature Mater. \textbf{3}, 793 (2004).
\bibitem{DomainWallSC4} W. Gillijns, A. Yu. Aladyshkin, M. Lange, M. J. Van Bael, and V. V. Moshchalkov, Phys. Rev. Lett. \textbf{95}, 227003 (2005).
\bibitem{DomainWallSC5} L. Y. Zhu, T. Y. Chen, and C. L. Chien, Phys. Rev. Lett. \textbf{101}, 017004 (2008).
\bibitem{SingleDomainWallJunctions1} A. F. Volkov and K. B. Efetov, Phys. Rev. B \textbf{78}, 024519 (2008).
\bibitem{SingleDomainWallJunctions2} A. I. Buzdin and A. S. Melnikov, Phys. Rev. B \textbf{67}, 020503 (2003).
\bibitem{TruptiPdNi2009} T. S. Khaire, W. P. Pratt, Jr., and N. O. Birge, Phys. Rev. B \textbf{79}, 094523 (2009).
\bibitem{SFISF1} Yu. S. Barash, I. V. Bobkova, and T. Kopp, Phys. Rev. B \textbf{66}, 140503(R) (2002).
\bibitem{SFISF2} T. Yu. Karminskaya, M. Yu. Kupriyanov, and A. A. Golubov, JETP Lett. \textbf{87}, 570 (2008).
\bibitem{SFISF3} V. N. Krivoruchko and E. A. Koshina, Phys. Rev. B \textbf{64}, 172511 (2001).
\bibitem{SFISF4} A. F. Volkov, F. S. Bergeret, and K. B. Efetov, Phys. Rev. Lett. \textbf{90}, 117006 (2003).
\bibitem{SFISF5} A. Vedyayev, C. Lacroix, N. Pugach and N. Ryzhanova, Europhys. Lett. \textbf{71}, 679 (2005).
\bibitem{SFISF6} C. Bell, G. Burnell, C. W. Leung, E. J. Tarte, D.-J. Kang, and M. G. Blamire, Appl. Phys. Lett. \textbf{84}, 1153(2004).
\bibitem{Zimansky} P. Cadden-Zimansky, Ya. B. Bazaliy, L. M. Litvak, J. S. Jiang, J. Pearson, J. Y. Gu, C.-Y. You, M. R. Beasley, and S. D. Bader, Phys. Rev. B \textbf{77}, 184501 (2008).
\bibitem{Parity} J. W. A. Robinson, G. B. Hal\'{a}sz, A. I. Buzdin and M. G. Blamire, arXiv:0808.0166 (unpublished).
\bibitem{Houzet2007} M. Houzet and A. I. Buzdin, Phys. Rev. B \textbf{76}, 060504(R) (2007).
\bibitem{Linder2009} J. Linder, T. Yokoyama, and P. Sudb{\o}, Phys. Rev. B \textbf{79}, 054523 (2009).
\bibitem{Sosnin2006} I. Sosnin, H. Cho, V. T. Petrashov, and A. F. Volkov, Phys. Rev. Lett. \textbf{96}, 157002 (2006).
\bibitem{Volkov2006} A. F. Volkov, A. Anishchanka, and K. B. Efetov, Phys. Rev. B \textbf{73}, 104412 (2006).
\bibitem{HalfMetal2} M. Eschrig and T. L\"{o}fwander, Nature Phys. \textbf{4}, 138 (2008).
\bibitem{HalfMetal3} M. Eschrig, J. Kopu, J. C. Cuevas, and G. Sch\"{o}n, Phys. Rev. Lett. \textbf{90}, 137003 (2003).
\bibitem{HalfMetal4} Y. Asano, Y. Sawa, Y. Tanaka, A. A. Golubov, Phys. Rev. B \textbf{76}, 224525 (2007).
\bibitem{HalfMetal5} A. V. Galaktionov, M. S. Kalenkov, A. D. Zaikin, Phys. Rev. B \textbf{77}, 094520 (2008).
\bibitem{HalfMetal1} R. S. Keizer, S. T. B. Goennenwein, T. M. Klapwijk, G. Miao, G. Xiao and A. Gupta, Nature (London) \textbf{439}, 825 (2006).
\bibitem{Ho1} W. C. Koehler, J. W. Cable, H. R. Child, M. K. Wilkinson, and E. O. Wollan, Phys. Rev. \textbf{158}, 450 (1967).
\bibitem{Ho2} S. Chikazumi, \emph{Physics of Ferromagnetism} (Oxford University Press, England, 1997).
\bibitem{Ho3} K. V. Rao , Phys. Rev. Lett. \textbf{22}, 943 (1969).
\bibitem{Ho4} J. Witt, S. Langridge, T. Hase, and M. G. Blamire (unpublished).
\bibitem{Ho5} J. Witt, J. W. A. Robinson, and M. G. Blamire (unpublished).
\bibitem{Bell2003} C. Bell, G. Burnell, D.-J. Kang, R. H. Hadfield, M. J. Kappers, and M. G. Blamire, Nanotechnology \textbf{14}, 630 (2003).
\end{references}
\end{document}